\documentclass[12pt,reqno]{amsart}

\usepackage{amsaddr}

\usepackage{amsmath,amsfonts,amsthm,amssymb}

\usepackage[a4paper,height=20cm,top=24mm,bottom=24mm,left=26mm,right=26mm]{geometry}

\usepackage[numbers,sort]{natbib}
\usepackage[breaklinks=true]{hyperref}

\usepackage[]{graphicx}            
\usepackage{booktabs}
\usepackage[table]{xcolor}

\DeclareMathAlphabet{\mathpzc}{OT1}{pzc}{m}{it}
\usepackage{titlesec}
\titleformat{\section}
{\bfseries\scshape\centering}
{\thesection.}{.5em}{}
\titleformat{\subsection}
{\rmfamily\bfseries}
{\thesubsection}{.5em}{}
\titleformat{\subsubsection}
{\rmfamily\bfseries}
{\thesubsubsection}{.5em}{}
\numberwithin{equation}{section}

\newcommand{\I}{\mathrm{i}}
\newcommand{\E}{\mathrm{e}}

\DeclareMathOperator{\ch}{ch}

\DeclareMathDelimiter{\Norm}{\mathord}{largesymbols}{"3E}{largesymbols}{"3E}

\DeclareMathOperator{\mult}{mult}
\DeclareMathOperator{\ord}{ord}
\begin{document}
\baselineskip 16pt
\parskip 8pt
\sloppy


\title{$\mathcal{N}=2$ Moonshine}


\author[T. Eguchi]{Tohru \textsc{Eguchi}}

\address{Department of Physics and Research Center for Mathematical Physics,
  Rikkyo University,
  Tokyo, Japan.}

\address{
  Yukawa Institute for 
  Theoretical Physics, 
  Kyoto University, 
  Kyoto, Japan.}

\email{tohru.eguchi@gmail.com}

\author[K. Hikami]{Kazuhiro \textsc{Hikami}}

\address{Faculty of Mathematics,
  Kyushu University,
  Fukuoka 819-0395, Japan.}

\email{khikami@gmail.com}


\date{September 3, 2012}

\begin{abstract}
We construct a model of moonshine phenomenon based on the use of $\mathcal{N}=2$ superconformal algebra. 
We consider an extremal Jacobi form of weight~$0$ and index~$2$,
and expand it in terms of $\mathcal {N}=2$ massless and massive representations. We find the multiplicities of massive representations are decomposed into a sum of dimensions of irreducible representations of the group $L_2(11)$.

\end{abstract}





\maketitle

\section{Introduction}
Study of the elliptic genus in string compactification by use of the
superconformal algebras (SCA) was introduced in~\cite{EguOogTaoYan89a}. 
In this approach we use the representation theory of SCA,
and decompose elliptic genus in terms of characters of SCA.  
In superconformal algebras  
there appear BPS (massless)  and non-BPS (massive) representations,
and massless characters are mock theta functions which possess unusual   
modular transformation
laws~\cite{EgucTaor88a,EgucTaor88b,EguchiHikami08a}.
Intrinsic structure of mock theta functions is revealed
in~\cite{Zweg02Thesis}
(see also~\cite{Zagier08a}).

Recently a phenomenon similar to the famous Monstrous
moonshine~\cite{ConwayNorton79} was discovered in this
analysis~\cite{EgucOoguTach10a}:
it was found that the 
expansion coefficients of $K3$ elliptic genus in terms of characters
of  $\mathcal{N}=4$  SCA are decomposed into a sum of dimensions of 
irreducible representations of Mathieu group $M_{24}$.
Analogues of McKay--Thompson series in the Monstrous moonshine were
constructed~\cite{MCheng10a,GabeHoheVolp10a,GabeHoheVolp10b,EguchiHikami10b},
and the decompositions into $M_{24}$ representations have been
verified up to very high degrees.
This phenomenon, sometimes called Mathieu moonshine, combines 
(mock) modular forms, a sporadic discrete group, and geometry of $K3$ surface 
in a curious manner.
Since $M_{24}$ is the symmetry group of an error correcting code
(Golay code)~\cite{Conway98a}, 
such a moonshine phenomenon may be  also interesting from the point of
view of a possible mechanism of information processing inside black holes.

Quite recently, a generalization of the Mathieu moonshine has been
proposed in~\cite{ChenDuncHarv12a}: authors of~\cite{ChenDuncHarv12a}
consider a sequence of higher dimensional analogues of 
Mathieu moonshine parametrized by $m=2,3,4,5,7$ where $(m-1)|24$:
$m=2$ case
corresponds to $K3$ and the original $M_{24}$  moonshine.
$m=3$ corresponds to a $4$-dimensional complex manifold.
From a general theory of Jacobi forms~\cite{EichZagi85}
it is known that the elliptic genera of complex $D$-dimensional
manifold are given by weak Jacobi forms of weight~$0$ and index~$D/2$.
When $D=4$, there exist 
two independent weak Jacobi forms with weight 0, index 2
\begin{align}
  &
  Z_1(z;\tau)
  =
  48
  \left[
    \left({\theta_{10}(z;\tau)\over
        \theta_{10}(0;\tau)}\right)^4+\left({\theta_{00}(z;\tau)\over
        \theta_{00}(0;\tau)}\right)^4+ \left({\theta_{01}(z;\tau)\over
        \theta_{01}(0;\tau)}\right)^4
  \right]  ,
  \\[2mm]
  &
  Z(z;\tau)
  =
  4
  \left[
    \left({\theta_{10}(z;\tau)\over \theta_{10}(0;\tau)}\cdot
      {\theta_{00}(z;\tau)\over \theta_{00}(0;\tau)}\right)^2
    +\left({\theta_{10}(z;\tau)\over \theta_{10}(0;\tau)}\cdot
      {\theta_{01}(z;\tau)\over \theta_{01}(0;\tau)}\right)^2
    +\left({\theta_{00}(z;\tau)\over \theta_{00}(0;\tau)}\cdot
      {\theta_{01}(z;\tau)\over \theta_{01}(0;\tau)}\right)^2
  \right] .
\end{align}
A suitable linear combination of these Jacobi forms,
$Z_1+n \, Z$,  
will reproduce an elliptic genus of some $4$-dimensional complex
manifold
($n = 15$ gives the elliptic genus of Hilbert scheme of two points on
$K3$ surface $K3^{[2]}$).
Coefficient of $Z_1$ is fixed since it contains the identity
representation in the NS sector.

Somewhat surprisingly authors of~\cite{ChenDuncHarv12a} chose to drop
$Z_1$ and studied $Z$ in isolation.
Decomposition of  $Z$ in terms of $\mathcal{N}=4$ characters  was
known in the literature~\cite{EguchiHikami09a}, 
and it was possible to guess  a new moonshine
phenomenon which is based on the Mathieu group $M_{12}$.
At higher
values of $m$ they apply a similar construction.
Drop the analogue of
$Z_1$ and consider a linear combination of other Jacobi forms which
possesses a polar term only in the massive representations of the
smallest isospin (an extremal Jacobi form)~\cite{ChenDuncHarv12a}. 
Thus in these examples we seem to lose the connection to geometry and
elliptic genus, however,  there still appear interesting new examples
of moonshine phenomena
($Z$ may still describe an elliptic genus of a non-compact
manifold~\cite{EgucSugaTaor07a}).

Ordinarily, $\mathcal{N}=4$ (resp. $\mathcal{N}=2$)
SCA describes the geometry of hyper-K{\"a}hler (resp. Calabi--Yau, CY
for short)
manifolds.
When one drops $Z_1$, however, it is not quite clear whether
$\mathcal{N}=4$ or
$\mathcal{N}=2$ is the relevant symmetry of the theory.   
In this article we take up the above example $Z$ at $D=4$,
and decompose it in terms of $\mathcal{N}=2$ SCA
characters~\cite{EguchiHikami10a} instead of $\mathcal{N}=4$~\cite{EguchiHikami09b}.
This is to see if it is possible to obtain further examples of moonshine phenomena. 
We in fact find a moonshine phenomenon with respect to the
group~$L_2(11)$ which is closely related to $M_{12}$.

\section{$\mathcal{N}=2$ Superconformal Algebras and Character Decomposition}

First let us recall the data of representation theory of $\mathcal{N}=2$ algebra. 
Representations of the extended $\mathcal{N}=2$ algebra with
central charge $c=3D$ were studied in~\cite{Odake90a,Odake90b}.
The characters of the extended algebra are obtained by summing over the spectral flow of irreducible $\mathcal{N}=2$ characters.  

There exist BPS (massless) and non-BPS (massive) representations in
the theory, parametrized by the  conformal weight $h$ and $U(1)$
charge $Q$.
In the Ramond sector $\widetilde{R}$ (with $(-1)^F$ insertion) characters are given as follows.
\begin{itemize}
\item massive (non-BPS) representations:\\
 $h> \frac{D}{8}$;
  $Q=\frac{D}{2}, \frac{D}{2} -1, \dots,-(\frac{D}{2}-1),
  -\frac{D}{2}$ and $Q\neq 0 \,(D=\mbox{even}) $,
  \begin{multline}
    \label{massive_character}
    \ch_{D,h,Q > 0}^{\widetilde{R},\mathcal{N}=2}(z;\tau)
    =
    (-1)^{Q+ \frac{D}{2} - 1} \, q^{h - \frac{D}{8}} \,
    \frac{
      \I \, \theta_{11}(z;\tau)}{
      \left[ \eta(\tau) \right]^3
    } \,
    \E^{2 \pi \I \left( Q - \frac{1}{2} \right) z}
    \\
    \times
    \sum_{n \in \mathbb{Z}}
    q^{\frac{D-1}{2} n^2 + \left( Q - \frac{1}{2} \right) n} \,
    \left(
      - \E^{2 \pi \I z}
    \right)^{(D-1) n} ,
  \end{multline}

\item massless (BPS) representations:\\
 $h=\frac{D}{8}$;
  $Q=\frac{D}{2}-1, \frac{D}{2}-2,
  \dots,
  - \left( \frac{D}{2} -1 \right)$,
  \begin{multline}
    \label{massless_character}
    \ch_{D,h=\frac{D}{8},Q\geq 0}^{\widetilde{R},\mathcal{N}=2}(z;\tau)
    =
    (-1)^{Q+ \frac{D}{2} } \,
    \frac{\I \, \theta_{11}(z;\tau)}{
      \left[ \eta(\tau) \right]^3} \,
    \E^{2 \pi \I \left( Q  + \frac{1}{2} \right) z }
    \\
    \times
    \sum_{n \in \mathbb{Z}}
    q^{\frac{D-1}{2} n^2 + \left( Q+\frac{1}{2} \right) n} \,
    \frac{
      \left( - \E^{2 \pi \I z} \right)^{(D-1) n}
    }{
      1 - \E^{2 \pi \I z} \, q^n
    } ,
  \end{multline}
  and for $h=\frac{D}{8}$; $Q=\frac{D}{2}$
  \begin{multline}
    \ch^{\widetilde{R},\mathcal{N}=2}_{D,h=\frac{D}{8},Q=\frac{D}{2}}(z;\tau)
    =
    (-1)^D \,
    \frac{
      \I \, \theta_{11}(z;\tau)}{
      \left[ \eta(\tau) \right]^3} \,
    \E^{2 \pi \I  \frac{D+1}{2} z}
    \\
    \times
    \sum_{n \in \mathbb{Z}}
    q^{\frac{D-1}{2} n^2 + \frac{D+1}{2} n}
    \,
    \frac{
      \left( 1-q \right) \,
      \left( 
        - \E^{2 \pi \I z}
      \right)^{(D-1) n} 
    }{
      \left( 1 - \E^{2 \pi \I z} \, q^n
      \right) \,
      \left( 1 - \E^{2 \pi \I z} \, q^{n+1}
      \right)
    } .
  \end{multline}
\end{itemize}
The characters for $Q<0$ are given by
\begin{equation}
  \ch_{D,h,- Q < 0}^{\widetilde{R},\mathcal{N}=2}(z;\tau)
  =
  \ch_{D,h,Q}^{\widetilde{R},\mathcal{N}=2}(-z ; \tau) .
\end{equation}

The Witten index of massless representations are given by
\begin{equation}
  \ch^{\widetilde{R},\mathcal{N}=2}_{D,h = \frac{D}{8}, Q \geq 0 }(z=0; \tau)
  =
  \begin{cases}
    (-1)^{Q+\frac{D}{2}} ,
    & \text{for $0 \leq Q < \frac{D}{2}$,}
    \\[2mm]
    1+ (-1)^D ,
    & \text{for $Q=\frac{D}{2}$,}
  \end{cases}
\end{equation}
while all massive representations having a vanishing index.

At the unitarity bound $h=\frac{D}{8}$, a massive character
decomposes into a sum of massless characters as
\begin{equation}
  \label{unitary_bound_Q}
  \lim_{h\searrow \frac{D}{8}} \ch_{D,h,Q+1}^{\widetilde{R},\mathcal{N}=2}(z;\tau)
  =
  \ch_{D,h=\frac{D}{8},Q+1}^{\widetilde{R},\mathcal{N}=2}(z;\tau)
  +
  \ch_{D,h=\frac{D}{8},Q}^{\widetilde{R},\mathcal{N}=2}(z;\tau)
  ,
\end{equation}
where $Q \geq 0$,
and
\begin{multline}
  \label{unitary_bound_zero}
  \lim_{h \searrow \frac{D}{8}}
    \ch_{D,h,Q=\frac{D}{2}}^{\widetilde{R},\mathcal{N}=2}(z;\tau)
  \\
  =
  \ch_{D, h=\frac{D}{8}, Q=\frac{D}{2}}^{\widetilde{R},\mathcal{N}=2}(z;\tau)
  +
  \ch_{D,h=\frac{D}{8},Q=\frac{D}{2}-1}^{\widetilde{R},\mathcal{N}=2}(z;\tau)
  +
  \ch_{D,h=\frac{D}{8},Q=- \left( \frac{D}{2}-1 \right)}^{\widetilde{R},\mathcal{N}=2}(z;\tau)
  .
\end{multline}

In our previous paper~\cite{EguchiHikami10a} we pointed out that when
the dimension~$D$ 
of CY manifold is odd, the 
decomposition of its elliptic genus into $\mathcal{N}=2$ characters 
becomes essentially the same as the decomposition of the 
elliptic genus  for a corresponding ($D-3$)-dimensional
hyper-K{\"a}hler manifold into $\mathcal{N}=4$ characters.
This is due to the 
uniqueness of Jacobi form of index~$3/2$ and weight~$0$.

In the case of even $D$,  however, the decomposition of CY manifolds
becomes somewhat different from that of hyper-K{\"a}hler manifolds.
For convenience we  introduce  
functions $B^{\mathcal{N}=2}_{D,Q}(z;\tau)$ and
$C^{\mathcal{N}=2}_D(z;\tau)$ by
\begin{gather}
  \begin{aligned}[b]
    B^{\mathcal{N}=2}_{D,Q}(z;\tau)
    & =
    (-1)^{Q+\frac{D}{2}-1} \,
    q^{
      -h + \frac{D}{8} + \frac{\left( Q - \frac{1}{2}\right)^2}{2(D-1)}}
    \,
    \left(
      \ch_{D,h> \frac{D}{8},Q}^{\widetilde{R},\mathcal{N}=2}(z;\tau)+
      \ch_{D,h> \frac{D}{8},-Q}^{\widetilde{R},\mathcal{N}=2}(z;\tau) 
    \right)
    \\
    & =
    \begin{cases}
      \displaystyle
      \frac{\I \, \theta_{11}(z;\tau)}{ \left[ \eta(\tau) \right]^3}
      \,
      \sum_{a=Q, D-Q}
      \sum_{n\in \mathbb{Z}}
       (-1)^n q^{
         \frac{1}{2(D-1)} 
         \left(
           (D-1)n+a-\frac{1}{2}
         \right)^2
       }
      \, \E^{
        2\pi \I  \left(  (D-1) n + a-\frac{1}{2} \right)  z
      }
      ,
      \\
      \hfill
      \text{for $ 1 \leq Q < \frac{D}{2}$,}
      \\[2mm]
      \displaystyle
      \frac{\I \, \theta_{11}(z;\tau)}{ \left[ \eta(\tau) \right]^3}
      \,
      \sum_{n\in \mathbb{Z}}
       (-1)^n q^{
         \frac{D-1}{2} 
         \left(
           n+\frac{1}{2}
         \right)^2
       }
      \, \E^{
        2\pi \I (D-1)  \left(   n + \frac{1}{2} \right) z}
      ,
      \\
      \hfill
      \text{for $Q=\frac{D}{2}$,}
    \end{cases}
  \end{aligned}
  \displaybreak[2]
  \\[2mm]
  \begin{aligned}[b]
    {C}^{\mathcal{N}=2}_D(z;\tau)
    & =
    (-1)^{\frac{D}{2}} \,
    \ch_{D, h=\frac{D}{8}, Q=0}^{\widetilde{R},\mathcal{N}=2}(z;\tau)
    \\
    & =
    \frac{\I \, \theta_{11}(z;\tau)}{
      \left[
        \eta(\tau)
      \right]^3
    } \,
    \E^{\pi \I z}
    \sum_{n \in \mathbb{Z}}
    (-1)^n \,
    q^{\frac{D-1}{2} n^2 + \frac{1}{2} n} \,
    \frac{
      \E^{2 \pi \I (D-1)n z}
    }{
      1 - \E^{2 \pi \I z} \, q^n
    } .
  \end{aligned}
\end{gather}  
$B_{D,Q}^{\mathcal{N}=2}$ stands for a charge $Q$ massive character
symmetrized under $z\leftrightarrow -z$. 
$C_{D}^{\mathcal{N}=2}$ is the massless character for the charge $Q=0$. 
The elliptic genus $Z_{CY_D}(z;\tau)$ for the Calabi--Yau $D$-fold (or
any weak Jacobi form of index~$D/2$ and weight~$0$)  is decomposed
as~\cite{EguchiHikami10a}
\begin{equation}
  \label{general_decomposition}
  Z_{CY_D}(z;\tau)
  =
 \chi \, C^{\mathcal{N}=2}_D(z;\tau)
  +
  \sum_{a=1}^{D/2}
  \Sigma_{D,a}(\tau) \, B^{\mathcal{N}=2}_{D,a}(z;\tau) .
\end{equation}
Here $\chi$ denotes the Euler number. 
From a 
mathematical point of view $\mathcal{N}=2$ decomposition~\eqref{general_decomposition}
gives 
a theta series expansion of 
(a real analytic) Jacobi form with a half-odd integral
index~\cite{EguchiHikami10a},
while $\mathcal{N}=4$ decomposition is that of a Jacobi
form of an integral index.
See also~\cite{DabhMurtZagi12a} for recent studies of Jacobi forms.
Since the massless character $C_D(z;\tau)$ is a mock theta function,
the generating functions
$\Sigma_{D,a}(\tau)$ for the multiplicity of massive representations become also mock
theta functions as far as $\chi \neq 0$.

In the case of $D=2$, 
the Calabi--Yau $2$-fold is the $K3$ surface.
The character
decomposition above 
reduces to the Mathieu moonshine considered in~\cite{EgucOoguTach10a}


\section{Moonshine from $\mathcal{N}=2$}

Let us now turn to the case $D=4$, and study a Jacobi form with
weight~$0$ and index~$2$,
\begin{equation}
  \begin{aligned}[b]
    &Z(z;\tau)
    \\
    &=
    4 \,
    \left[
      \left(
        \frac{\theta_{10}(z;\tau)}{\theta_{10}(0;\tau)} \cdot
        \frac{\theta_{00}(z;\tau)}{\theta_{00}(0;\tau)}
      \right)^2
      +
      \left(
        \frac{\theta_{00}(z;\tau)}{\theta_{00}(0;\tau)} \cdot
        \frac{\theta_{01}(z;\tau)}{\theta_{01}(0;\tau)}
      \right)^2
      +
      \left(
        \frac{\theta_{01}(z;\tau)}{\theta_{01}(0;\tau)} \cdot
        \frac{\theta_{10}(z;\tau)}{\theta_{10}(0;\tau)}
      \right)^2
    \right] 
    \\
    &
    =
    \frac{1}{12} \, \left[ \phi_{0,1}(z;\tau) \right]^2
    -
    \frac{1}{12} \, E_4(\tau) \,
    \left[ \phi_{-2,1}(z;\tau) \right]^2.
  \end{aligned}
  \label{genus_1A}
\end{equation}
As computed in~\cite{EguchiHikami10a}, we have a decomposition 
\begin{eqnarray}
  Z(z;\tau)
 =&&\hskip-5mm 
  12 \, C^{\mathcal{N}=2}_4(z;\tau)\nonumber 
  +
  \\
&&  \hskip-18mm + q^{-\frac{1}{24}} \,  \left(
    -2 + 10 \, q + 20 \, q^2 + 42 \, q^3 + 62 \, q^4 + 118 \, q^5 +170\,q^6+270\,q^7+
    \cdots
  \right) \, B^{\mathcal{N}=2}_{4,1}(z;\tau)\nonumber
  \label{decompose_1A}
  \\
 && \hskip-18mm + q^{-\frac{3}{8}} \,  \left(
    12 \, q + 36 \, q^2 + 60 \, q^3 + 120 \, q^4 + 180 \, q^5 +312\,q^6+456\,q^7+
    \cdots
  \right) \, B^{\mathcal{N}=2}_{4,2}(z;\tau)\nonumber \\
  &&\label{expand}
  \\
  =&&\hskip-5mm
  8 \ch^{\widetilde{R},\mathcal{N}=2}_{D=4,h=\frac{1}{2},Q=0}(z;\tau)
  -2
  \ch^{\widetilde{R},\mathcal{N}=2}_{D=4,h=\frac{1}{2},Q=1}(z;\tau)
  -2
  \ch^{\widetilde{R},\mathcal{N}=2}_{D=4,h=\frac{1}{2},Q=-1}(z;\tau)\nonumber 
  \\
 &&\hskip-5mm  -
  \sum_{Q=\pm 1 , \pm 2}
  (-1)^Q
  \sum_{n=1}^\infty
  p_{|Q|}(n) \,
  \ch^{\widetilde{R},\mathcal{N}=2}_{D=4,h=n+\frac{1}{2},Q}(z;\tau) .
\end{eqnarray}
Expansion coefficients of the massive representations in~\eqref{expand}  
suggest the group
$L_2(11)$ being relevant for a moonshine phenomenon.
$L_2(11)$ is the group
$PSL_2(\mathbb{F}_{11})$ of $2\times 2$ matrices of determinant one with
matrix elements in the field $\mathbb{F}_{11}$~\cite{Conway98a,ATLAS85}.

See Table~\ref{tab:table_SL2_11} for a character table of
$SL_2(11)\cong 2.L_2(11)$,
which is a  double cover of $L_2(11)$~\cite{ATLAS85}.
Therein $n_g$ denotes the number of elements in conjugacy class $g$,
and the orthogonality relation reads as
\begin{equation}
  \label{character_orthogonal}
  \sum_g n_g \, \chi_R^g \, \overline{\chi_{R^\prime}^g} =
  |G| \, \delta_{R, R^\prime} .
\end{equation}
$|G|$ denotes the order of $G$,
and 
$\left| SL_2(11) \right|=2^3\cdot 3 \cdot 5 \cdot 11=1320$.
It is easy to check that at small values of $n$, the number of massive 
representations $p_a(n)$ can be written as a sum of dimensions of irreducible
representations $R$ of $SL_2(11)$
\begin{equation}
  \sum_{R} \mult_{R,a}(n) \, \dim R = p_a(n) ,
\end{equation}
with multiplicities $\mult_{R,a}(n)$
\begin{gather*}
  10=5+5, \hskip2mm 20=2\times 10, \hskip2mm 42=10+10+2\times 11, \cdots\\
   12=6+6, \hskip2mm 36=6+6+12+12,\hskip2mm 60=3\times (6+6) + 12+12,
   \cdots
 \end{gather*}


\begin{table}[htbp]
  \newcolumntype{L}{>{$}l<{$}}
  \newcolumntype{R}{>{$}r<{$}}
  \newcolumntype{C}{>{$}c<{$}}
  \rowcolors{2}{gray!16}{}
  \centering
  \resizebox{.97\textwidth}{!}{
    \begin{tabular}[]{C*{15}{C}}
      \toprule
      n_g &
      1 & 1 & 110 & 132 & 132 & 132 & 132 & 60 & 60 & 110 & 110 & 60 &
      60 & 110 & 110
      \\
      \midrule
      R \backslash g&
      \mathrm{1A}&
      \mathrm{2A} & \mathrm{4A}& \mathrm{5A}& \mathrm{5B}&
      \mathrm{10A}& \mathrm{10B}&
      \mathrm{11A}& \mathrm{11B}&
      \mathrm{12A}& \mathrm{12B}&
      \mathrm{22A} & \mathrm{22B} &
      \mathrm{3A}&  \mathrm{6A} 
      \\
      \midrule\midrule
      {\chi_1} &
      1 & 1 & 1&  1&   1&   1&  1&  1&  1&  1&  1&  1&  1&   1&  1
      \\
      {\chi_2} &
      5& 5&1&0&0&0&0&\frac{-1+\I\sqrt{11}}{2}&\frac{-1-\I\sqrt{11}}{2}&1&1&\frac{-1-\I\sqrt{11}}{2}&\frac{-1+\I\sqrt{11}}{2}&-1&-1
      \\
      {\chi_3} &
      5&5&1&0&0&0&0&\frac{-1-\I\sqrt{11}}{2}&\frac{-1+\I\sqrt{11}}{2}&1&1&\frac{-1+\I\sqrt{11}}{2}&\frac{-1-\I\sqrt{11}}{2}&-1&-1
      \\
      {\chi_4} &
      10&10&-2&0&0&0&0&-1&-1&1&1&-1&-1&1&1
      \\
      {\chi_5} &
      10&10&2&0&0&0&0&-1&-1&-1&-1&-1&-1&1&1
      \\
      {\chi_6} &
      11&11&-1&1&1&1&1&0&0&-1&-1&0&0&-1&-1
      \\
      {\chi_7} &
      12&12&0&\frac{-1-\sqrt{5}}{2}&\frac{-1+\sqrt{5}}{2}&\frac{-1+\sqrt{5}}{2}&\frac{-1-\sqrt{5}}{2}&1&1&0&0&1&1&0&0
      \\
      {\chi_8} &
      12&12&0&\frac{-1+\sqrt{5}}{2}&\frac{-1-\sqrt{5}}{2}&\frac{-1-\sqrt{5}}{2}&\frac{-1+\sqrt{5}}{2}&1&1&0&0&1&1&0&0
      \\
      \midrule
      {\chi_9} &
      6&-6&0&1&1&-1&-1&\frac{1-\I\sqrt{11}}{2}&\frac{1+\I\sqrt{11}}{2}&0&0&\frac{-1-\I\sqrt{11}}{2}&\frac{-1+\I\sqrt{11}}{2}&0&0
      \\
      {\chi_{10}} &
      6&-6&0&1&1&-1&-1&\frac{1+\I\sqrt{11}}{2}&\frac{1-\I\sqrt{11}}{2}&0&0&\frac{-1+\I\sqrt{11}}{2}&\frac{-1-\I\sqrt{11}}{2}&0&0
      \\
      {\chi_{11}} &
      10&-10&0&0&0&0&0&-1&-1&0&0&1&1&-2&2
      \\
      {\chi_{12}} &
      10&-10&0&0&0&0&0&-1&-1&-\sqrt{3}&\sqrt{3}&1&1&1&-1
      \\
      {\chi_{13}} &
      10&-10&0&0&0&0&0&-1&-1&\sqrt{3}&-\sqrt{3}&1&1&1&-1
      \\
      {\chi_{14}} &
      12&-12&0&\frac{-1+\sqrt{5}}{2}&\frac{-1-\sqrt{5}}{2}&\frac{1+\sqrt{5}}{2}&\frac{1-\sqrt{5}}{2}&1&1&0&0&-1&-1&0&0
      \\
      {\chi_{15}} &
      12&-12&0&\frac{-1-\sqrt{5}}{2}&\frac{-1+\sqrt{5}}{2}&\frac{1-\sqrt{5}}{2}&\frac{1+\sqrt{5}}{2}&1&1&0&0&-1&-1&0&0
      \\
      \bottomrule
    \end{tabular}
  }
  \caption{Character table for $SL_2(11) \cong 2. L_2(11)$~\cite{ATLAS85}.}
  \label{tab:table_SL2_11}
\end{table}
  
It is known that $L_2(11)$ has a permutation representation on
$12$~symbols
(see, \emph{e.g.},~\cite{Conway98a}).
Representatives of conjugacy classes $g$ are given in
Table~\ref{tab:represent_conjugacy}.

\begin{table}[htbp]
  \newcolumntype{L}{>{$}l<{$}}
  \newcolumntype{R}{>{$}r<{$}}
  \newcolumntype{C}{>{$}c<{$}}
  \rowcolors{2}{gray!16}{}
  \centering
  \begin{tabular}[]{CLL}
    \toprule
    g & \text{cycle shape}&
    \text{permutation}
    \\
    \midrule
    \mathrm{1A} &
    1^{12} &
    ()
    \\
    \mathrm{5A} &
    1^2 5^2 &
    (3,5,7,9,11)(4,6,8,10,12)
    \\
    \mathrm{5B} &
    1^2 5^2 &
    (3,7,11,5,9)(4,8,12,6,10)    
    \\
    \mathrm{11A} &
    1^1 11^1 &
    (2,3,4,11,5,7,12,10,6,9,8)
    \\
    \mathrm{11B} &
    1^1 11^1 &
    (2,4,5,12,6,8,3,11,7,10,9)
    \\
    \mathrm{4A} &
    2^6 &
    (1,2)(3,4)(5,12)(6,11)(7,10)(8,9)
    \\
    \mathrm{3A} &
    3^4 &
    (1,2,3)(4,8,12)(5,10,9)(6,11,7)
    \\
    \mathrm{12AB} &
    6^2 &
    (1,2,3,10,4,11)(5,6,12,8,9,7)
    \\
    \bottomrule    
  \end{tabular}
  \caption{Permutation representatives of conjugacy classes of
    $L_2(11)$.
    Note that the name of conjugacy class
    is for that of $SL_2(11)$.
  }
  \label{tab:represent_conjugacy}
\end{table}

As in the case of Mathieu moonshine, we want to construct twisted
elliptic genus $Z_g$  for 
each conjugacy class $g$.
It turns out that due to complication of double covering of the group
$L_2(11)$
we can not construct twisted elliptic genera for all classes.
However,
in the following 
we   obtain 
those twisted elliptic genera which are just enough to determine the
decomposition of the multiplicities of massive representations into
the sum of irreducible representations of $SL_2(11)$.

Let us call the representations $\{\chi_i\}$,
$i=1, 2, 3, 4, 5, 6, 7, 8$ in Table~\ref{tab:table_SL2_11} as even and representations 
$\{\chi_j\}$,
$j=9, 10, 11, 12, 13, 14, 15$ as odd, respectively.
We assume as in~\cite{ChenDuncHarv12a} that
multiplicities of 
$|Q|= 1$ massive representations are decomposed into a sum of even
representations,
and that    those of $|Q|=2$ massive representations are 
decomposed into a sum of odd representations.

Twisted elliptic genus is a Jacobi form with weight~$0$ and index~$2$ and 
has a decomposition analogous to~\eqref{decompose_1A},\begin{equation}
  Z_g(z;\tau) 
  =
  \chi_g \, C^{\mathcal{N}=2}_4(z;\tau)
  +
  \Sigma_{g,1}(\tau) \, B^{\mathcal{N}=2}_{4,1}(z;\tau)
  +
  \Sigma_{g,2}(\tau) \, B^{\mathcal{N}=2}_{4,2}(z;\tau) .\label{expansion_twisted_genus}
\end{equation}
Here $\chi_g$ is the Euler number,
$\chi_g=Z_g(0;\tau)$,
and
$\Sigma_{g,a}(\tau)$ are $q$-series with integral Fourier coefficients
\begin{equation}
  \Sigma_{g,a}(\tau) =
  q^{-\frac{(2a-1)^2}{24}}
  \sum_{n=0}^\infty p_{g,a}(n) \, q^n . \label{Sigma_g}
\end{equation}
Structure of the character table suggests that 
the conjugacy classes $\mathrm{5A}$ and
$\mathrm{5B}$ have the same twisted elliptic genus, and we use the 
notation $\mathrm{5AB}$.
Similarly we assume 
the same for classes $\mathrm{10A,B}$, $\mathrm{11A,B}$ and
$\mathrm{22A,B}$,
 and use the notations 
$\mathrm{10AB}$, $\mathrm{11AB}$, and $\mathrm{22AB}$, respectively.

In view of the character table and the permutation representatives of
conjugacy classes, we suppose
that the Euler number for class $g$ is given by
\begin{equation}
  \chi_g =
  {\chi}_{{1}}^g +
  {\chi}_{{6}}^g .
\end{equation}
We then find
\begin{equation*}
   \begin{array}{c*{6}{c}}
     \toprule
     ~~~g~~~
     &  \mathrm{1A} & \mathrm{5AB} & \mathrm{11AB} & \mathrm{4A}
     & \mathrm{3A}
     & \mathrm{12AB}
     \\
     \midrule
     \chi_g & 12 & 2 & 1 & 0 & 0 & 0
     \\
     \bottomrule
   \end{array}
 \end{equation*}
Thus the classes
$\{\mathrm{1A}, \mathrm{5AB}, \mathrm{11AB}\}$
belong to type~I and
$\{\mathrm{4A}, \mathrm{3A}, \mathrm{12AB}\}$ belong to type~II in the
terminology of~\cite{EguchiHikami10b}.

The original elliptic genus~\eqref{genus_1A}
is  for the class $g=\mathrm{1A}$.
By trial and error we have constructed the twisted elliptic genera
$Z_g(z;\tau)$
for classes $g=
\mathrm{5AB}$,
$\mathrm{11AB}$,
$\mathrm{4A}$,
$\mathrm{3A}$,
$\mathrm{12AB}$,
which are presented in Table~\ref{tab:twisted_genus_N2}.

\begin{table}[htbp]
  \newcolumntype{L}{>{$}l<{$}}
  \newcolumntype{R}{>{$}r<{$}}
  \newcolumntype{C}{>{$}c<{$}}
  \rowcolors{2}{gray!13}{}
  \centering
  \resizebox{.97\textwidth}{!}{
    \begin{tabular}{CL}
      \toprule
      g & Z_g^{\mathcal{N}=2}(z;\tau)
      \\
      \midrule
      \midrule
      \mathrm{1A} &
      \begin{aligned}[t]
         \frac{1}{12}
         \left[ \phi_{0,1}(z;\tau) \right]^2 -
         \frac{1}{12} E_4(\tau) \,
         \left[ \phi_{-2,1}(z;\tau) \right]^2
      \end{aligned}
      \\
      \midrule
      \mathrm{5AB}
      &
      \begin{aligned}[t]
        &
        \frac{1}{72} \left[ \phi_{0,1}(z;\tau) \right]^2
        \\
        & \quad
        +
        \left(
          -\frac{5}{576} \, \phi_2^{(3)}(\tau)
          +\frac{25}{288} \, \phi_2^{(5)}(\tau)
          +\frac{35}{576} \, \phi_2^{(15)}(\tau)
          +
          \frac{5}{16} \,
          \eta(\tau) \, \eta(3\tau) \, \eta(5\tau) \, \eta(15\tau)
        \right) \,
        \phi_{0,1}(z;\tau) \phi_{-2,1}(z;\tau)
        \\
        & \quad
        +
        \Biggl(
          -\frac{1}{192} E_4(\tau) 
          +
          \frac{25}{16} \, \left[ \eta(\tau) \, \eta(5\tau) \right]^4
          -
          \frac{75}{4} \, \left[
            \eta(\tau) \, \eta(3\tau) \, \eta(5\tau) \, \eta(15\tau)
          \right]^2
          \\
        & \qquad   \quad
          +\frac{5}{96} \, \left[ \phi_2^{(3)}(\tau) \right]^2
          -\frac{25}{576} \, \left[ \phi_2^{(5)}(\tau) \right]^2
          -\frac{5}{32} \, \phi_2^{(3)}(\tau) \, \phi_2^{(5)}(\tau)
          \\
          & \qquad \qquad
          +\frac{175}{96} \, \phi_2^{(3)}(\tau) \, \phi_2^{(15)}(\tau)
          -\frac{175}{96} \, \phi_2^{(5)}(\tau) \, \phi_2^{(15)}(\tau)
        \Biggr) \,
        \left[  \phi_{-2,1}(z;\tau) \right]^2
        \\
      \end{aligned}
      \\
      \midrule
      \mathrm{11AB} &
      \begin{aligned}[t]
        &
         \frac{1}{144} \left[ \phi_{0,1}(z;\tau) \right]^2
         +
         \left(
           \frac{11}{72} \phi_2^{(11)}(z;\tau)
           +
           \frac{11}{20} \left[ \eta(\tau) \eta(11\tau) \right]^2
         \right) \phi_{-2,1}(z;\tau) \phi_{0,1}(z;\tau)
         \\
         &
         +
         \left(
           \frac{1}{120} E_4(\tau)
           -\frac{121}{720} \left[ \phi_2^{(11)}(z;\tau) \right]^2
           + \frac{1089}{100} \phi_2^{(11)}(z;\tau) \,
           \left[ \eta(\tau) \eta(11\tau) \right]^2
           -
           \frac{121}{125}
           \left[ \eta(\tau) \eta(11\tau) \right]^4
         \right)
         \left[  \phi_{-2,1}(z;\tau) \right]^2
      \end{aligned}
      \\
      \midrule
      \mathrm{4A} &
      \begin{aligned}[t]
        -2 \,
        \frac{\eta(\tau) \, \eta(2\tau)}{\eta(4 \tau)} \,
        B^{\mathcal{N}=2}_{4,1}(z;\tau)
      \end{aligned}
      \\
      \midrule
      \mathrm{12AB} &
      \begin{aligned}[t]
        \left(
          \frac{\eta(\tau) \, \eta(2\tau)}{\eta(4\tau)}
          +3 \,
          \frac{
            \left[ \eta(3\tau) \right]^2 \, \eta(6\tau)
          }{
            \eta(\tau) \, \eta(12\tau)}
          -6 \,
          \frac{
            \eta(4\tau) \, \left[ \eta(6\tau) \right]^4
          }{
            \eta(2\tau) \, \eta(3\tau) \,
            \left[ \eta(12\tau) \right]^2
          }
        \right) \,
        B^{\mathcal{N}=2}_{4,1}(z;\tau)
      \end{aligned}
      \\
      \midrule
      \mathrm{3A}
      &
      \begin{aligned}[t]
        -2 \,
        \frac{ \left[ \eta(2\tau) \right]^3}{
          \eta(\tau) \, \eta(6\tau)} \,
        B^{\mathcal{N}=2}_{4,1}(z;\tau)
      \end{aligned}
      \\
      \bottomrule
    \end{tabular}
  }
  \caption{Twisted elliptic genus $Z_g(z;\tau)$.}
  \label{tab:twisted_genus_N2}
\end{table}

In the case of conjugacy class $\mathrm{2A}$ we assume
\begin{equation}
  \begin{gathered}
    \Sigma_{\mathrm{1A}, 1}(\tau) = \Sigma_{\mathrm{2A},  1}(\tau)  ,
    \\
    \Sigma_{\mathrm{1A}, 2}(\tau) = -\Sigma_{\mathrm{2A}, 2}(\tau)  ,
  \end{gathered}
  \label{sign_change_Sigma}
\end{equation}
corresponding to the sign change in the odd sector of character table
(see Table~\ref{tab:table_SL2_11}).
We suppose a similar pairing as above (sign change in the $\Sigma_{g,2}$ part) 
between
$\mathrm{5AB}$ and $\mathrm{10AB}$,
 $\mathrm{11AB}$ and $\mathrm{22AB}$.

In the case of $\mathrm{4A}$, on the other hand, 
we set
the odd part to vanish 
\begin{equation}
  \Sigma_{\mathrm{4A},2}(\tau)=0
  \label{vanish_Sigma}
\end{equation}
since the odd elements in the 
character table all vanish (see Table~\ref{tab:table_SL2_11}) for
class~$\mathrm{4A}$.
We also assume that the conjugacy classes,
$\mathrm{12AB}$, $\mathrm{3A}$, and $\mathrm{6A}$,
have vanishing odd parts.



In the case of the Mathieu moonshine, all the twisted elliptic genera were  
Jacobi forms on congruence subgroup $\Gamma_0(\ord(g))$ with a possible 
character. 
In the present case  only the twisted elliptic genera of conjugacy
classes $\mathrm{1A}, 
\mathrm{5AB},
\mathrm{11AB}, 
\mathrm{4A}, 
\mathrm{12AB},
\mathrm{3A}=\mathrm{6A}$   
are Jacobi forms 
(level of congruence subgroup is sometimes higher than
$\ord(g)$).
Due to the sign flip in odd sector~\eqref{sign_change_Sigma} 
twisted elliptic genera of the other classes can not be Jacobi
forms. 
If we insist that twisted elliptic genera must be Jacobi forms, 
twisted elliptic genera do not exist for classes $\mathrm{2A}$,
$\mathrm{10AB}$,
$\mathrm{22AB}$. 
This situation is similar to the $\mathcal{N}=4$ moonshine in~\cite{ChenDuncHarv12a}.

In
Table~\ref{tab:massive_N2}, the Fourier coefficients of
$\Sigma_{g,a}(\tau)$,
\emph{i.e.}, the number of the massive representations are given.
We have omitted from Table
the odd sector $p_{g,2}$ for  classes $g$
whose  generating
functions vanish identically 
$\Sigma_{g,2}(\tau)=0$.


\begin{table}[htbp]
    \newcolumntype{L}{>{$}l<{$}}
    \newcolumntype{R}{>{$}r<{$}}
    \newcolumntype{C}{>{$}c<{$}}
    \rowcolors{2}{gray!13}{}
    \centering
  \resizebox{.9\textwidth}{!}{
    \begin{tabular}[]{Rc*{3}{R}*{3}{R}c*{3}{R}}
    \toprule
    &\phantom{ab}&
    \multicolumn{6}{c}{$p_{g,1}(n)$} &
    \phantom{ab} & \multicolumn{3}{c}{$p_{g,2}(n)$}
    \\
    \cmidrule{3-8}
    \cmidrule{10-12}
    n \backslash g
    &&
    \mathrm{1A} &\mathrm{5AB} &\mathrm{11AB}& \mathrm{4A}&
    \mathrm{12AB} & \mathrm{3A}
    &&
    \mathrm{1A} &\mathrm{5AB} &\mathrm{11AB}
    \\
    \midrule
 0 & &  -2  &  -2  &  -2  &  -2  &  -2  &  -2  &&  0  &  0  &  0
\\
  1  &&  10  &  0  &  -1  &  2  &  2  &  -2  &&  12  &  2 &   1
\\
  2  &&  20  &  0  &  -2  &  4  &  -2  &  2  &&  36  &  1  &  3
\\
  3  & & 42  &  2  &  -2  &  -2  &  -2  &  0  &&  
  60  &  5  &  5
\\ 
 4 & &  62  &  2  &  -4  &  -2  &  4  &  2  & & 120  &  10  &  -1
\\
  5  &&    118  &  -2  &  -3  &  -2  &  4  &  4 & &  180  &  15  &  4
\\
  6  &&  170  &  0  &  -6  &  2  &  8  &  -4  & & 312  &    27  &  4
\\
  7  &&  270  &  0  &  -5  &  -2  &  4  &  0  & & 456  &  36  &  5
\\
  8  & & 400  &  0  &  -7  &  -4  &    14  &  4  &&  720  &  60  &  5
\\
  9  &&  600  &  0  &  -5  &  4  &  10  &  0  & & 1020  &  85  &  8
\\
  10 & &    828  &  -2  &  -8  &  4  &  10  &  0 & &  1524  &  129  &  6
  \\
  11  &&  1220  &  0  &  -12  &  0  &  12  &  2  &&    2124  &  179  &  12
\\ 
 12  &&  1670  &  0  &  -13  &  -2  &  22  &  -4 & &  3036  &  251  & 11
\\
  13  &&    2330  &  0  &  -13  &  2  &  14  &  -4  &&  4140  &  345  &  15
\\
  14 & &  3162  &  2  &  -17  &  2  &  20  &    6  &&  5760  &  480  &  18
\\ 
 15  &&  4316  &  -4  &  -18  &  0  &  30  &  -4  &&  7740  &  645  &    18
\\
  16  &&  5730  &  0  &  -23  &  -6  &  42  &  0  &&  10512  &  877  &  18
\\ 
 17  &&  7710  &    0  &  -23  &  2  &  38  &  6  &&  13896  &  1156  & 25
\\ 
18  &&  10102  &  2  &  -29  &  6  &  42  &  -8  &&    18540  &  1545  &
27
\\
  19  &&  13312  &  2  &  -31  &  -4  &  50  &  -2  &&  24240  &  2020  &    29
\\
  20  &&  17298  &  -2  &  -38  &  -6  &  66  &  6  &&  31824  &  2654  &  34
\\
  21  &&  22500  &    0  &  -39  &  0  &  72  &  -6  &&  41124  &  3429  &  39
\\
  22  &&  28860  &  0  &  -48  &  4  &  70  &  0  &&    53292  &  4437
  &  41
\\ 
 23  &&  37162  &  2  &  -51  &  -2  &  82  &  10  &&  68220  &  5685  &
 53
\\
  24  &&  47262  &  2  &  -60  &  -6  &  96  &  -12 & &  87420  &  7285
  &  58
\\  
25  &&    60128  &  -2  &  -64  &  4  &  112  &  -4  &&  110880  &  9240&  66
\\
  26  &&  75900  &  0  &  -77  &    8  &  116  &  12  &&  140724  &
11729  &  67
\\
  27  &&  95740  &  0  &  -81  &  -4  &  128  &  -8  &&    177072  &
14752  &  82
\\
  28  &&  119860  &  0  &  -95  &  -4  &  152  &  -2  &&  222780  &
18565 &   85
\\
  29  &&  150062  &  2  &  -99  &  2  &  170  &  14  &&  278280  &  23190  &  101
\\
  30 & &    186576  &  -4  &  -116  &  8  &  182  &  -12 & &  347424  &  28954  &  110
\\
  31  &&  231800  &    0  &  -124  &  -4  &  206  &  -4  &&  431136  &  35931  &  123
\\
  32  &&  286530  &    0  &  -141  &  -10  &  236  &  18  &&  534492  &
  44537  &  134
\\ 
 33  &&  353694  &  4  &  -154  &  6 &   252  &  -12  &&  659220  &
 54935  &  155
\\
  34  &&  434524  &  4  &  -174  &  12  &  270  &  -2  &&    812160  &
  67680  &  162
\\
  35  &&  533334  &  -6  &  -188  &  -2  &  310  &  18  &&  996084  &
  83009  &  188
\\
  36  &&  651790  &  0  &  -213  &  -10  &  350  &  -20  &&  1220124  &
  101679  &    202
\\
  37  &&  795490  &  0  &  -228  &  2  &  380  &  -8  &&  1488612  &
  124047  &  224
\\
  38  &&    967490  &  0  &  -257  &  10  &  400  &  26  &&  1813860
&  151155  &  246
\\
  39  &&  1174962  &    2  &  -278  &  -6  &  450  &  -18 & &  2202420
&  183535  &  275
\\
  40  &&    1422264  &  -6  &  -311  &  -12  &  504  &  0  &&  2670564
&  222549  &  292
\\
  41  & & 1719450 &   0  &  -334  &  6  &  546  &  30  &&  3228048  &
269008  &  329
\\
  42  &&  2072480  &  0  &  -371  &    12  &  588  &  -28 & &  3896568
&  324708  &  357
\\
  43  &&  2494542  &  2  &  -401  &  -6  &    648  &  -6  &&  4690320
&  390860  &  393
\\
  44  &&  2994874  &  4  &  -448  &  -14  &  718  &  34 &&   5637960  &
  469830  &  427
\\ 
 45  &&  3590404  &  -6  &  -480  &  4  &  778  &  -26  &&    6759744  &
 563314  &  475
\\ 
 46  &&  4294020  &  0  &  -534  &  12  &  834  &  -6  &&  8093748  &
 674483  &  509
\\
  47  &&  5128880  &  0  &  -574  &  -4  &  908  &  38  &&  9668448  &
  805698  &    570
\\
  48  &&  6112362  &  2  &  -635  &  -18  &  1002  &  -36 & &  11534040
  &  961170  &    606
\\ 
 49  &&  7274774  &  4  &  -681  &  10  &  1090  &  -10  &&  13730220  &
 1144185  &    669
\\
  50  &&  8641024  &  -6  &  -752  &  20  &  1166  &  46  &&  16323228
  &  1360273  &  724
\\
    \bottomrule
  \end{tabular}
  }
  \caption{The number of massive representations,
    $p_{g,1}(n)$ and $p_{g,2}(n)$}
  \label{tab:massive_N2}
\end{table}

In order to test the moonshine conjecture,
we have computed multiplicities
$\mult_{R,a}(n)$ of representations $R$
\begin{equation}
  p_{g,a}(n)
  =
  \sum_{R} \mult_{R,a}(n) \,
  {\chi}_{R}^g  .
  \label{p_and_multiplicity}
\end{equation}
Here $R$ runs over irreducible representations from
${\chi_1}$
to ${\chi_8}$
(resp.
from ${\chi_9}$ to ${\chi_{15}}$)
for $a=1$
(resp. $a=2$).
From the orthogonality relation~\eqref{character_orthogonal}, we have
\begin{equation}
  \mult_{R,a}(n)
  =
  \sum_g \frac{n_g}{|G|} \,
  \overline{\chi_R^g} \, p_{g,a}(n). 
\end{equation}
See Table~\ref{tab:multiplicity_N2} for the results of the decomposition into 
irreducible representations. 
We find that
the multiplicities 
$\mult_{R,1}(n)$ are the same for $R=\chi_2$ and $R=\chi_3$,
and also for
$R=\chi_7$ and
$R=\chi_8$ in the even sector.
In the odd sector we have
$\chi_9=\chi_{10}$,
$\chi_{11}=\chi_{12}=\chi_{13}$, and $\chi_{14}=\chi_{15}$.

We have verified up to $n= 100$
the positivity and integrality of the
multiplicities $\mult_{R,a}(n)$,
and consider this to be a strong evidence for a $\mathcal{N}=2$ moonshine.


\begin{table}[htbp]
  \newcolumntype{L}{>{$}l<{$}}
  \newcolumntype{R}{>{$}r<{$}}
  \newcolumntype{C}{>{$}c<{$}}
  \rowcolors{2}{gray!13}{}
  \centering
  \resizebox{.97\textwidth}{!}{
  \begin{tabular}[]{Rc*{6}{R}c*{3}{R}}
    \toprule
    & \phantom{a} &
  \multicolumn{6}{c}{$\mult_{R,1}(n)$} &
  \phantom{abc} &
  \multicolumn{3}{c}{$\mult_{R,2}(n)$}
  \\
      \cmidrule{3-8}
    \cmidrule{10-12}
    n \backslash R
    && {\chi_1} &
    {\chi_{2}}
    =
    {\chi_3}
    & {\chi_4} &
    {\chi_5} &
    {\chi_6} &
    {\chi_{7}}
    =
    {\chi_{8}}
    &&
      {\chi_{9}}
      =
      {\chi_{10}}
      &
      {
        \begin{array}{r}
          {\chi_{11}}
          =
          {\chi_{12}}
          \\
          =
          {\chi_{13}}
        \end{array}
      }
    &
      {\chi_{14}}
      =
      {\chi_{15}}
    \\
    \midrule
 0&&
 -2& 0& 0& 0& 0& 0&& 0& 0& 0
\\
   1
&& 0& 1& 0& 0& 0& 0&& 1& 0& 0
\\
   2
&&   0& 0& 0& 2& 0& 0&& 1& 0& 1
\\
   3
&& 0& 0& 1& 1& 2& 0&& 3& 0& 1
\\
   4
&& 1&   1& 3& 1& 1& 0&& 5& 2& 0
\\
   5
&& 0& 1& 4& 2& 0& 2&& 8& 2& 1
\\ 
  6
&& 0& 4&   4& 2& 2& 2&& 14& 4& 1
\\
   7
&& 0& 3& 6& 4& 4& 4&& 19& 6& 2
\\
   8
&& 2& 5&   11& 5& 4& 6&& 31& 10& 2
\\
   9
&& 2& 7& 11& 9& 8& 10&& 44& 14& 3
\\
   10
&& 1&   9& 15& 13& 11& 14&& 66& 22& 3
\\
   11
&& 2& 12& 23& 19& 18& 20&& 92& 30&   5
\\
   12
&& 3& 18& 31& 23& 25& 28& &129& 44& 7
\\
   13
&& 3& 22& 39& 35& 37&   40&& 177& 60& 9
\\
   14
&& 7& 28& 55& 49& 49& 54&& 246& 84& 12
\\
   15
&& 6&   40& 73& 63& 66& 76&& 330& 114& 15
\\
   16
&& 11& 52& 99& 83& 89& 100&&   448& 156& 19
\\
   17
&& 15& 66& 128& 116& 121& 136&& 591& 206& 26
\\
   18
&&   17& 88& 163& 151& 163& 178&& 789& 276& 33
\\
   19
&& 23& 112& 216& 198&   215& 236&& 1031& 362& 42
\\
   20
&& 30& 144& 282& 258& 276& 308&& 1354&   476& 54
\\
   21
&& 38& 187& 359& 335& 364& 402&& 1749& 616& 69
\\
   22
&& 47&   235& 457& 435& 469& 516&& 2263& 800& 89
\\
   23
&& 63& 298& 588& 560&   605& 666&& 2899& 1024& 113
\\
   24
&& 75& 381& 742& 708& 775& 848&& 3714&   1314& 143
\\
   25
&& 97& 481& 940& 904& 983& 1082&& 4710& 1668&   180
\\
   26
&& 123& 600& 1184& 1148& 1243& 1366&& 5977& 2120& 225
\\
   27
&&   150& 755& 1486& 1442& 1576& 1726&& 7518& 2668& 284
\\
   28
&& 189& 942&   1859& 1807& 1973& 2162&& 9459& 3360& 353
\\
   29
&& 241& 1172& 2322&   2266& 2471& 2710&& 11815& 4198& 440
\\
   30
&& 289& 1457& 2875& 2817&   3079& 3372&& 14750& 5244& 546
\\
   31
&& 362& 1802& 3569& 3499& 3830&   4192&& 18303& 6510& 675
\\
   32
&& 450& 2219& 4411& 4329& 4734& 5184& &  22686& 8074& 835
\\
   33
&& 550& 2738& 5426& 5344& 5856& 6402&& 27981&   9960& 1027
\\
   34
&& 674& 3354& 6658& 6572& 7198& 7868&& 34470& 12276&   1260
\\
   35
&& 826& 4106& 8170& 8066& 8832& 9664&& 42276& 15058&   1543
\\
   36
&& 1003& 5018& 9971& 9851& 10809& 11812&& 51782& 18450&   1885
\\ 
  37
&& 1226& 6112& 12156& 12030& 13196& 14422&& 63172& 22514&   2297
\\
   38
&& 1491& 7416& 14775& 14645& 16053& 17544&& 76974& 27438&   2793
\\
   39
&& 1802& 9004& 17926& 17774& 19512& 21312&& 93461& 33320&   3387
\\
   40
&& 2179& 10886& 21692& 21520& 23619& 25804&& 113324& 40410&   4099
\\
   41
&& 2641& 13143& 26208& 26028& 28561& 31202&& 136979& 48850&   4950
\\
   42
&& 3167& 15838& 31560& 31368& 34447& 37614&& 165339& 58974&   5970
\\
   43
&& 3814& 19043& 37977& 37759& 41470& 45282&& 199019& 70994&   7178
\\
   44
&& 4582& 22842& 45586& 45342& 49792& 54370&& 239225& 85346&   8620
\\
   45
&& 5476& 27378& 54612& 54354& 59712& 65194&& 286821& 102334&   10328
\\ 
  46
&& 6548& 32720& 65294& 65020& 71428& 77976&& 343419&   122540& 12355
\\
   47
&& 7824& 39052& 77973& 77669& 85324& 93148& &  410226& 146388&
   14754
\\
   48
&& 9306& 46535& 92891& 92551& 101714&   111018&& 489378& 174648&
17586
\\
   49
&& 11081& 55358& 110526& 110166&   121067& 132144&& 582555& 207912& 20925
\\
   50
&& 13157& 65719& 131260&   130878& 143811& 156974&& 692568& 247190&
   24863
\\
    \bottomrule
  \end{tabular}
  }
  \caption{
    Multiplicities $\mult_{R,a}(n)$ up to $n=50$.
    }
  \label{tab:multiplicity_N2}
\end{table}


\section{Discussions}

In this paper we have taken up the suggestion of~\cite{ChenDuncHarv12a} on the extremal Jacobi form. We 
studied the decomposition of an extremal  
form of index 2 into characters of $\mathcal{N}=2$ SCA. 
We have found a strong evidence for a moonshine phenomenon 
with respect to the  group $L_2(11)$,
which is a subgroup or $M_{12}$ which appeared
in~\cite{ChenDuncHarv12a} from decomposition into $\mathcal{N}=4$ SCA characters.

Currently, however, the real origin of the moonshine phenomenon is not very well understood and still remains rather mysterious. 
It seems that we have to 
construct and study more examples of moonshine phenomena before we figure out the workings behind them. 
Especially the 
$\mathcal{N}=2$  decomposition of models of~\cite{ChenDuncHarv12a} for
higher values of $m=4,5,7$ may be 
good candidates of  moonshine with the group $L_2(7),L_2(5),L_2(3)$, respectively.



\section*{Acknowledgments}
This work is supported in part by Grant-in-Aid 
\#23340115, \#22224001, \#24654041, \#22540069 from the Ministry of
Education, Culture, Sports, Science and Technology of Japan.

\clearpage

\begin{center}
\textbf{Appendix}
\end{center}
\appendix
\section{Modular Forms}
As usual we set $q=\E^{2 \pi \I \tau}$ where $\tau$ is in the upper
half-plane.
The Dedekind $\eta$-function is
\begin{equation}
  \eta(\tau) = 
  q^{\frac{1}{24}} \, \prod_{n=1}^\infty
  \left( 1 - q^n \right) .
\end{equation}
The Eisenstein series $E_{2k}(\tau)$ is
\begin{equation}
  E_{2k}(\tau)
  =
  1 - \frac{4 \, k}{B_{2k}}
  \sum_{n=1}^\infty 
  \left(
    \sum_{1 \leq r | n} r^{2k-1}
  \right) \, q^n ,
\end{equation}
where $B_k$ is the Bernoulli number
\begin{equation*}
  \frac{t}{\E^t -1}
  = \sum_{k=0}^\infty B_k \, \frac{t^k}{k!}  .
\end{equation*}
We use modular form of weight 2 on $\Gamma_0(M)$
\begin{equation}
  \phi_2^{(M)}(\tau)
  =
  \frac{24}{M-1} q \frac{\partial}{\partial q}
  \log \frac{\eta(M\tau)}{\eta(\tau)} .
\end{equation}

The Jacobi theta functions are defined as
\begin{equation}
  \begin{aligned}
    \theta_{11}(z;\tau)
    & =
    \sum_{n \in \mathbb{Z}}
    q^{\frac{1}{2} \left( n+ \frac{1}{2} \right)^2} \,
    \E^{2 \pi \I \left(n+\frac{1}{2} \right) \,
      \left( z+\frac{1}{2} \right)
    }
    ,
    \\[2mm]
    \theta_{10}(z;\tau)
    & =
    \sum_{n \in \mathbb{Z}}
    q^{\frac{1}{2} \left( n + \frac{1}{2} \right)^2} \,
    \E^{2 \pi \I \left( n+\frac{1}{2} \right) z}
    ,
    \\[2mm]
    \theta_{00} (z;\tau)
    & =
    \sum_{n \in \mathbb{Z}}
    q^{\frac{1}{2} n^2} \,
    \E^{2 \pi \I  n  z}
    ,
    \\[2mm]
    \theta_{01} (z;\tau)
    & =
    \sum_{n \in \mathbb{Z}}
    q^{\frac{1}{2} n^2} \,
    \E^{2 \pi \I n \left( z+\frac{1}{2} \right) }
    .
  \end{aligned}
\end{equation}
Some of the Jacobi forms are given by use of these $q$-series as
\begin{align}
  \phi_{-2,1}(z;\tau)
  & =
  - \frac{
    \left[ \theta_{11}(z;\tau) \right]^2}{
    \left[ \eta(\tau) \right]^6
  } ,
  \\[2mm]
  \phi_{0,1}(z;\tau)
  & =
  4 \,
  \left[
    \left(
      \frac{\theta_{10}(z;\tau)}{\theta_{10}(0;\tau)}
    \right)^2
    +
    \left(
      \frac{\theta_{00}(z;\tau)}{\theta_{00}(0;\tau)}
    \right)^2
    +
    \left(
      \frac{\theta_{01}(z;\tau)}{\theta_{01}(0;\tau)}
    \right)^2
  \right]  .
  \label{phi01}
\end{align}
See~\cite{EichZagi85} for general properties of the Jacobi forms.

\clearpage 
\section{$\mathcal{N}=4 $ Moonshine}

In order to compare the decomposition in $\mathcal{N}=2$ and
$\mathcal{N}=4$ SCA,
we reproduce the analysis of the $m=3$ case
in~\cite{ChenDuncHarv12a} 
where the $M_{12}$ moonshine is observed.
See Table~\ref{tab:character_2M12} for the character table of
$2.M_{12}$, which is a double cover of $M_{12}$. 
The order is
$|2.M_{12}|=
2^7 \cdot 3^3 \cdot 5\cdot 11
=190080$.


\begin{table}[h]
  \newcolumntype{L}{>{$}l<{$}}
  \newcolumntype{R}{>{$}r<{$}}
  \newcolumntype{C}{>{$}c<{$}}
  \rowcolors{2}{gray!13}{}
  \centering
  \rotatebox[]{90}{
    \resizebox{.93\textheight}{!}{
      \begin{tabular}[]{C*{26}{C}}
        \toprule
        n_g &
        1& 1& 792& 495& 495& 1760& 1760& 2640& 2640& 5940& 5940
        & 9504& 9504 &15840& 15840& 15840& 11880& 11880& 11880&
        11880& 9504& 9504& 8640 &8640& 8640& 8640
        \\
        \midrule
        R \backslash g
        &
        \mathrm{1A}&                \mathrm{2A}&
        \mathrm{4A}&
        \mathrm{2B}&                 \mathrm{2C}& 
        \mathrm{3A} & \mathrm{6A}& \mathrm{3B}& \mathrm{6B}&
        \mathrm{4B} &
        \mathrm{4C}&                 \mathrm{5A}&                 \mathrm{10A}& 
        \mathrm{12A}& \mathrm{6C}&                \mathrm{6D}&
        \mathrm{8A}& \mathrm{8B}&
        \mathrm{8C} & \mathrm{8D} &\mathrm{20A}&\mathrm{20B}
        &\mathrm{11A}&\mathrm{22A}&\mathrm{11B}&\mathrm{22B}
        \\
        \midrule \midrule
        \chi_1 &
        1 &1 &1 &1 &1 &1 &1 &1 &1 &1 &1 &1 &1 &1 &1 &1 &1 &1
        &1 &1 &1&1&1&1&1&1
        \\
        \chi_2 &
        11& 11&-1&3&3&2&2&-1&-1&-1&3&1 &1&-1&0&0&-1&-1&1&1
        &-1&-1&0&0&0&0
        \\
        \chi_3 &
        11& 11&-1&3&3&2&2&-1&-1&3&-1&1 &1&-1&0&0&1&1&-1&-1
        &-1&-1&0&0&0&0
        \\
        \chi_4 &
        16& 16&4&0&0&-2&-2&1&1&0&0&1 &1 &1&0&0&0&0&0&0
        &-1&-1&\frac{-1+\I\sqrt{11}}{2}&\frac{-1+\I\sqrt{11}}{2}&\frac{-1-\I\sqrt{11}}{2}&\frac{-1-\I\sqrt{11}}{2}
        \\
        \chi_5 &
        16& 16&4&0&0&-2&-2&1&1&0&0&1 &1& 1&0&0&0&0&0&0
        &-1&-1&\frac{-1-\I\sqrt{11}}{2}&\frac{-1-\I\sqrt{11}}{2}&\frac{-1+\I\sqrt{11}}{2}&\frac{-1+\I\sqrt{11}}{2}
        \\
        \chi_6 &
        45 &45&5&-3&-3&0&0&3&3&1&1&0 &0&-1&0&0&-1&-1&-1&-1
        &0&0&1&1&1&1
        \\
        \chi_7 &
        54 &54&6&6&6&0&0&0&0&2&2&-1&-1 &0&0&0&0&0&0&0
        &1&1&-1&-1&-1&-1
        \\
        \chi_8 &
        55 &55&-5&7&7&1&1&1&1&-1&-1&0 &0& 1&1&1&-1&-1&-1&-1
        &0&0&0&0&0&0
        \\
        \chi_9 &
        55 &55&-5&-1&-1&1&1&1&1&3&-1&0 &0& 1&-1&-1&-1&-1&1&1
        &0&0&0&0&0&0
        \\
        \chi_{10} &
        55 &55&-5&-1&-1&1&1&1&1&-1&3&0 &0 &1&-1&-1&1&1&-1&-1
        &0&0&0&0&0&0
        \\
        \chi_{11} &
        66 &66&6&2&2&3&3&0&0&-2&-2&1 &1& 0&-1&-1&0&0&0&0
        &1&1&0&0&0&0
        \\
        \chi_{12} &
        99 &99&-1&3&3&0&0&3&3&-1&-1&-1&-1&-1&0&0&1&1&1&1
        &-1&-1&0&0&0&0
        \\
        \chi_{13} &
        120&120&0&-8&-8&3&3&0&0&0&0&0 &0 &0&1&1&0&0&0&0
        &0&0&-1&-1&-1&-1
        \\
        \chi_{14} &
        144&144&4&0&0&0&0&-3&-3&0&0&-1&-1 &1&0&0&0&0&0&0
        &-1&-1&1&1&1&1
        \\
        \chi_{15} &
        176&176&-4&0&0&-4&-4&-1&-1&0&0&1 &1&-1&0&0&0&0&0&0
        &1&1&0&0&0&0
        \\
        \midrule
        \chi_{16} &
        10&-10&0&-2&2&1&-1&-2&2&0&0&0 &0 &0&1&-1&\I \sqrt{2}&-\I \sqrt{2}&\I \sqrt{2}&-\I \sqrt{2}
        &0&0&-1&1&-1&1
        \\
        \chi_{17} &
        10&-10&0&-2&2&1&-1&-2&2&0&0&0 &0 &0&1&-1&-\I \sqrt{2}&\I \sqrt{2}&-\I \sqrt{2}&\I \sqrt{2}
        &0&0&-1&1&-1&1
        \\
        \chi_{18} &
        12&-12&0&4&-4&3&-3&0&0&0&0&2&-2 &0&1&-1&0&0&0&0
        &0&0&1&-1&1&-1
        \\
        \chi_{19} &
        32&-32&0&0&0&-4&4&2&-2&0&0&2&-2 &0&0&0&0&0&0&0
        &0&0&-1&1&-1&1
        \\
        \chi_{20} &
        44&-44&0&4&-4&-1&1&2&-2&0&0&-1 &1& 0&1&-1&0&0&0&0
        &\I \sqrt{5}&-\I \sqrt{5}&0&0&0&0
        \\
        \chi_{21} &
        44&-44&0&4&-4&-1&1&2&-2&0&0&-1 &1 &0&1&-1&0&0&0&0
        &-\I \sqrt{5}&\I \sqrt{5}&0&0&0&0
        \\
        \chi_{22} &
        110&-110&0&-6&6&2&-2&2&-2&0&0&0 &0& 0&0&0&\I \sqrt{2}&-\I \sqrt{2}&-\I \sqrt{2}&\I \sqrt{2}
        &0&0&0&0&0&0
        \\
        \chi_{23} &
        110&-110&0&-6&6&2&-2&2&-2&0&0&0 &0& 0&0&0&-\I \sqrt{2}&\I \sqrt{2}&\I \sqrt{2}&-\I \sqrt{2}
        &0&0&0&0&0&0
        \\
        \chi_{24} &
        120&-120&0&8&-8&3&-3&0&0&0&0&0 &0& 0&-1&1&0&0&0&0
        &0&0&-1&1&-1&1
        \\
        \chi_{25} &
        160&-160&0&0&0&-2&2&-2&2&0&0&0 &0 &0&0&0&0&0&0&0
        &0&0&\frac{1-\I\sqrt{11}}{2}&\frac{-1+\I\sqrt{11}}{2}&\frac{1+\I\sqrt{11}}{2}&\frac{-1-\I\sqrt{11}}{2}
        \\
        \chi_{26} &
        160&-160&0&0&0&-2&2&-2&2&0&0&0& 0& 0&0&0&0&0&0&0
        &0&0&\frac{1+\I\sqrt{11}}{2}&\frac{-1-\I\sqrt{11}}{2}&\frac{1-\I\sqrt{11}}{2}&\frac{-1+\I\sqrt{11}}{2}
        \\
        \bottomrule
      \end{tabular}                
    }}
  \caption{Character table of $2.M_{12}$~\cite{ATLAS85}.}
  \label{tab:character_2M12}
\end{table}

Twisted elliptic genus $Z_g^{\mathcal{N}=4}(z;\tau)$ of $\mathcal{N}=4$ theory
are summarized in Table~\ref{tab:twisted_genus}. 
As in the case of $\mathcal{N}=2$, not all the conjugacy classes have
the elliptic genus.
Namely,  
when $g$ and $g^\prime$  is a pair of classes  with the same values of characters in the 
even sector ($\chi_i^g=\chi_i^{g'}$,
for
$i=1,\cdots, 15$) and
opposite values in the odd sector
($\chi_i^g=-\chi_i^{g'}$,
for
$i=16,\cdots, 26$), only either $Z_g$ or $Z_{g'}$ is 
 a Jacobi form and becomes a twisted genus.

Note that
the $\mathcal{N}=2$ twisted elliptic genus for type-I classes are 
somewhat similar to those of $\mathcal{N}=4$, and  
we have $Z_{g}^{\mathcal{N}=2}(z;\tau)=Z_g^{\mathcal{N}=4}(z;\tau)$ for
$g=\mathrm{1A}$,~$\mathrm{11AB}$,
and
\begin{equation}
  Z_{\mathrm{5AB}}^{\mathcal{N}=2}(z;\tau)
  =
  Z_{\mathrm{5A}}^{\mathcal{N}=4}(z;\tau)
  + 
  5 \, \frac{
    \left[ \eta(15 \tau) \right]^3}{
    \eta(\tau) \, \eta(5 \tau)
  } \,
  B^{\mathcal{N}=2}_{4,2}(z;\tau) .
\end{equation}


\begin{table}[h]
  \newcolumntype{L}{>{$}l<{$}}
  \newcolumntype{R}{>{$}r<{$}}
  \newcolumntype{C}{>{$}c<{$}}
  \rowcolors{2}{gray!13}{}
  \centering
  \resizebox{.97\textwidth}{!}{
    \begin{tabular}{CL}
      \toprule
      g &  Z_g^{\mathcal{N}=4}(z;\tau)
      \\
      \midrule
      \midrule
      \mathrm{1A} &
      \begin{aligned}[t]
     &
        \frac{1}{12}
        \left[ \phi_{0,1}(z;\tau) \right]^2 -
        \frac{1}{12} E_4(\tau) \,
        \left[ \phi_{-2,1}(z;\tau) \right]^2
        \\
      \end{aligned}
      \\
      \midrule
      \mathrm{2B} &
      \begin{aligned}[t]
        &
        \frac{1}{36} \, \left[\phi_{0,1}(z;\tau) \right]^2 
         +\frac{1}{9} \, \phi_2^{(2)}(\tau) \, \phi_{0,1}(z;\tau) \,
        \phi_{-2,1}(z;\tau)
        +
        \left(
          - \frac{5}{36} \, E_4(\tau) 
          + \frac{128}{3} \, \frac{ \left[ \eta(2 \tau) \right]^{16}}{
            \left[ \eta(\tau) \right]^8}
        \right)
        \left[ \phi_{-2,1}(z;\tau) \right]^2
        \\
      \end{aligned}
      \\
      \midrule
      \mathrm{3A} &
      \begin{aligned}[t]
        &
        \frac{1}{48} \left[ \phi_{0,1}(z;\tau) \right]^2
        + \frac{1}{8} \phi_2^{(3)}(\tau) \, \phi_{0,1}(z;\tau) \,
        \phi_{-2,1}(z;\tau)
        +
        \left(
          \frac{11}{48} \left[ \phi_2^{(3)}(\tau) \right]^2
          - \frac{3}{8} E_4(3\tau)
        \right) \, \left[ \phi_{-2,1}(z;\tau) \right]^2
      \end{aligned}
      \\
      \midrule
      \mathrm{5A}
      &
      \begin{aligned}[t]
        &
        \frac{1}{72} \left[ \phi_{0,1}(z;\tau) \right]^2
        +
        \frac{5}{36} \phi_2^{(5)}(\tau) \phi_{0,1}(z;\tau) \phi_{-2,1}(z;\tau)
        \\
        &
        +
        \left(
          \frac{1}{48} E_4(\tau) 
          - \frac{25}{144}  \left[ \phi_2^{(5)}(\tau) \right]^2
          +\frac{25}{4} \left[ \eta(\tau) \eta(5\tau) \right]^4
        \right) \left[
          \phi_{-2,1}(z;\tau) \right]^2
        \\
      \end{aligned}
      \\
      \midrule
      \mathrm{6C} &
      \begin{aligned}[t]
        &
        \frac{1}{144} \left[ \phi_{0,1}(z;\tau) \right]^2
        +
        \left(
          \frac{5}{24} \phi_2^{(6)}(\tau) 
          -\frac{1}{24} \phi_2^{(3)}(\tau) 
          - \frac{1}{72} \phi_2^{(2)}(\tau) 
        \right) \phi_{0,1}(z;\tau) \phi_{-2,1}(z;\tau)
        \\
        &
        +
        \left(
          \frac{1}{36} \left[ \phi_2^{(2)}(\tau) \right]^2
          +       \frac{1}{16} \left[ \phi_2^{(3)}(\tau) \right]^2
          + \frac{5}{4} \phi_2^{(2)}(\tau) \, \phi_2^{(6)}(\tau)
          -\frac{5}{4} \phi_2^{(3)}(\tau) \,  \phi_2^{(6)}(\tau)
          -\frac{1}{4} \phi_2^{(2)}(\tau) \, \phi_2^{(3)}(\tau)
        \right) \left[
          \phi_{-2,1}(z;\tau) \right]^2
      \end{aligned}
      \\
      \midrule
      \mathrm{8C} &
      \begin{aligned}[t]
        &
        \frac{1}{72} \left[ \phi_{0,1}(z;\tau) \right]^2
        +
        \left(
          \frac{5}{72} \phi_2^{(2)}(\tau) 
          -\frac{1}{8} \phi_2^{(4)}(\tau) 
          + \frac{7}{36} \phi_2^{(8)}(\tau) 
        \right) \phi_{0,1}(z;\tau) \phi_{-2,1}(z;\tau)
        \\
        &
        +
        \left(
          \frac{1}{4} \left[ \phi_2^{(4)}(\tau) \right]^2
          - \frac{5}{24} \phi_2^{(2)}(\tau) \, \phi_2^{(4)}(\tau)
          +\frac{35}{36} \phi_2^{(2)}(\tau) \,    \phi_2^{(8)}(\tau)
          -\frac{7}{6} \phi_2^{(4)}(\tau) \, \phi_2^{(8)}(\tau)
        \right) \left[
          \phi_{-2,1}(z;\tau) \right]^2
      \end{aligned}
      \\
      \midrule
      \mathrm{11AB} &
      \begin{aligned}[t]
      &   \frac{1}{144} \left[ \phi_{0,1}(z;\tau) \right]^2
         +
         \left(
           \frac{11}{72} \phi_2^{(11)}(z;\tau)
           +
           \frac{11}{20} \left[ \eta(\tau) \eta(11\tau) \right]^2
         \right) \phi_{-2,1}(z;\tau) \phi_{0,1}(z;\tau)
         \\
         &
         +
         \left(
           \frac{1}{120} E_4(\tau)
           -\frac{121}{720} \left[ \phi_2^{(11)}(z;\tau) \right]^2
           + \frac{1089}{100} \phi_2^{(11)}(z;\tau) \,
           \left[ \eta(\tau) \eta(11\tau) \right]^2
           -
           \frac{121}{125}
           \left[ \eta(\tau) \eta(11\tau) \right]^4
         \right)
         \left[  \phi_{-2,1}(z;\tau) \right]^2
      \end{aligned}
      \\
      \midrule 
      \mathrm{4A} &
      \begin{aligned}[t]
        -2 \,
        \frac{\left[ \eta(\tau) \right]^4}{\left[ \eta(2 \tau)     \right]^3}
        B_2^{(1),\,\mathcal{N}=4}(z;\tau)
      \end{aligned}
      \\
      \midrule
      \mathrm{3B} &
      \begin{aligned}[t]
        &
        \left( \frac{1}{6} \phi_2^{(9)}(\tau) 
          - \frac{9}{2}
          \frac{\left[\eta(9\tau)\right]^6}{\left[\eta(3\tau)\right]^2}
        \right) \, \phi_{-2,1}(z;\tau) \, \phi_{0,1}(z;\tau)
        +
        \left(
          \frac{27}{2} \left[ \eta(3\tau) \right]^8
          -
          \frac{1}{6} \frac{ \left[\eta(\tau)
            \right]^{12}}{\left[\eta(3\tau) \right]^4}
        \right) \,
        \left[ \phi_{-2,1}(z;\tau) \right]^2
      \end{aligned}
      \\
      \midrule
      \mathrm{4B} &
    \begin{aligned}[t]
        &
        -2 \,
        \frac{\left[ \eta(2\tau) \right]^3}{\left[ \eta(4 \tau)
          \right]^2}
        B_2^{(1),\,\mathcal{N}=4}(z;\tau)
        \\
        & =
        \left(
          \frac{1}{36} \phi_2^{(2)}(\tau) 
          -\frac{1}{4} \phi_2^{(4)}(\tau) 
          + \frac{7}{18} \phi_2^{(8)}(\tau) 
        \right) \phi_{0,1}(z;\tau) \phi_{-2,1}(z;\tau)
        \\
        &
        +
        \left(
          \frac{5}{36} \left[ \phi_2^{(2)}(z;\tau) \right]^2
          +  \frac{3}{2} \left[ \phi_2^{(4)}(z;\tau) \right]^2
          - \frac{17}{12} \phi_2^{(2)}(z;\tau) \, \phi_2^{(4)}(z;\tau)
          +\frac{35}{18} \phi_2^{(2)}(z;\tau) \,    \phi_2^{(8)}(z;\tau)
          -\frac{7}{3} \phi_2^{(4)}(z;\tau) \, \phi_2^{(8)}(z;\tau)
        \right) \left[
          \phi_{-2,1}(z;\tau) \right]^2
      \end{aligned}
      \\
      \midrule
      \mathrm{12A} &
      \begin{aligned}[t]
        -2 \,
        \frac{ \eta(\tau) \, \eta(3\tau)}{ \eta(6 \tau) } \,
        B_2^{(1),\,\mathcal{N}=4}(z;\tau)
      \end{aligned}
      \\
      \midrule
      \mathrm{8A} &
      \begin{aligned}[t]
        -2 \,
        \frac{ \left[ \eta(4\tau) \right]^4}{ \eta(2 \tau)  \left[
            \eta(8\tau)\right]^2 } \,
        B_2^{(1),\,\mathcal{N}=4}(z;\tau)
      \end{aligned}
      \\
      \midrule
      \mathrm{20A} &
      \begin{aligned}[t]
        -2 \,
        \frac{ \left[\eta(2 \tau) \right]^2  \eta(5\tau)}{
          \eta(\tau) \, \eta(10 \tau) } \,
        B_2^{(1),\,\mathcal{N}=4}(z;\tau)
      \end{aligned}
      \\
      \bottomrule
    \end{tabular}
  }
  \caption{Twisted elliptic genus for $\mathcal{N}=4$ SCA.}
  \label{tab:twisted_genus}
\end{table}

In $\mathcal{N}=4$ SCA,
twisted  elliptic genera in Table~\ref{tab:twisted_genus} are
decomposed as~\cite{EguchiHikami08a,EguchiHikami09b}
\begin{equation}
  Z_g^{\mathcal{N}=4}(z;\tau)
  =
  \chi_g \,
  \ch^{\widetilde{R},\,\mathcal{N}=4}_{k=2,h=\frac{2}{4}, \ell=0}(z;\tau)
  +
  \Sigma_g^{(1)}(\tau) \, B_{2}^{(1),\,\mathcal{N}=4}(z;\tau)
  +
  \Sigma_g^{(2)}(\tau) \, B_{2}^{(2),\,\mathcal{N}=4}(z;\tau) ,
  \label{N=4decompo}
\end{equation}
where
$\chi_g$ is the Witten index, 
$\chi_g = Z^{\mathcal{N}=4}_g(z=0;\tau)$,
and is given by $\chi_g=\chi_1^g+\chi_2^g$,
\begin{equation*}
  {  \begin{array}{c*{12}{c}}
    \toprule
    g &
    \mathrm{1A} & \mathrm{2B} & \mathrm{3A}
    & \mathrm{5A} & \mathrm{6C} & \mathrm{8C} &
    \mathrm{11AB} &\text{others}
    \\
    \midrule
    \chi_g &
    12 & 4 & 3 
    & 2 & 1 & 2 & 1 & 0
    \\
    \bottomrule
  \end{array}}
\end{equation*}
$\mathcal{N}=4$ massless characters
and bases of  massive characters are respectively
given as
\begin{gather}
  \ch^{\widetilde{R},\,\mathcal{N}=4}_{k=2,h=\frac{2}{4}, \ell=0}(z;\tau)
  =
  {\theta_{11}(z;\tau)^2\over \eta(\tau)^3}{\I \over
    \theta_{11}(2z;\tau)}\sum_{n\in {\mathbb{Z}}}
  q^{3n^2} \,  
  \E^{12 \pi \I n z}\,{1+q^n \E^{2\pi \I z}\over 1-q^n \E^{2 \pi \I z}},
\\
B_2^{(a),\,\mathcal{N}=4}(z;\tau)={\theta_{11}(z;\tau)^2\over
  \eta(\tau)^3} \chi_{1,{a-1\over 2}}(z;\tau) ,
\end{gather}
where $\chi_{1,j}$ is an $SU(2)$ spin $j$ affine character at level $1$.

The $q$-series $\Sigma_g^{(a)}(\tau)$ is the generating function of
the number of $\mathcal{N}=4$ massive representations, and we have
\begin{gather}
  \Sigma_g^{(a)}(\tau) = q^{-\frac{a^2}{12}}
  \sum_{n=0}^\infty A_g^{(a)}(n) \, q^n .
\end{gather}
For comparison with our $\mathcal{N}=2$ moonshine,
values of the Fourier coefficients $A_{g}^{(a)}(n)$ are given in 
Tables~\ref{tab:decompose_twisted}.
Note that as in the case of $\mathcal{N}=2$
the sign change of odd part in the character table is reflected in \emph{e.g.}
$\Sigma_{\mathrm{2A}}^{(2)}(\tau)=
-\Sigma_{\mathrm{1A}}^{(2)}(\tau)$.
It should be remarked that
we have 
$\Sigma_{\mathrm{4C}}^{(1)}(\tau) = \Sigma_{\mathrm{2B}}^{(1)}(\tau)$.

Multiplicities of massive representations $A_g^{(a)}(n)$ are given by formula
like~\eqref{p_and_multiplicity} with the character table for
$2.M_{12}$ in Table~\ref{tab:character_2M12}. 
Multiplicities of irreducible representations are completely determined
by Table~\ref{tab:decompose_twisted}.

\begin{table}[htbp]
  \newcolumntype{L}{>{$}l<{$}}
  \newcolumntype{R}{>{$}r<{$}}
  \newcolumntype{C}{>{$}c<{$}}
  \rowcolors{2}{gray!13}{}
  \centering
  \rotatebox[]{90}{
  \resizebox{.93\textheight}{!}{
  \begin{tabular}[]{Rc*{7}{R}*{6}{R}c*{8}{R}}
    \toprule
    & \phantom{ab}
    &
    \multicolumn{13}{C}{A_{g}^{(1)}(n)}
    & \phantom{ab}
    &
    \multicolumn{8}{C}{A_{g}^{(2)}(n)}
    \\
    \cmidrule{3-15}
    \cmidrule{17-24}
    n \backslash g
    & & \mathrm{1A} &\mathrm{2B} &\mathrm{3A}& 
    \mathrm{5A} & \mathrm{6C} & \mathrm{8C} &
    \mathrm{11AB}
    & \mathrm{4A}& \mathrm{3B}& \mathrm{4B}& \mathrm{12A}&
    \mathrm{8A}& \mathrm{20A}
    & &
    \mathrm{1A} &\mathrm{2B} &\mathrm{3A}& 
    \mathrm{5A} & \mathrm{6C} & \mathrm{8C} &
    \mathrm{11AB}
    & \mathrm{3B}
    \\
    \midrule
    0 && -2  & -2 & -2 & -2 &
    -2 & -2 & -2 & 
    -2 &-2 &-2 &-2 &-2 &-2 
    && 0  & 0 & 0 & 0 &  0 & 0 & 0 &  0
    \\
    1 &&
    32 & 0 & -4 
 &   2 & 0 & 0 & -1 &
    8 &2 & 0 & 2 &0 &-2
    &&    20 & -4 & 2   &   0 & 2 & -2 & -2 &    -4
    \\
    2 &&
    110 &  -2 & 2 
 &    0 & -2 &2 & 0 &
    -10 & 2 &  6& 2 & -2 & 0
    &&    88 &  8 & -2  &   -2  & 2 &4 & 0 &    4
    \\
    3 &&
    288 & 0 & 0
 &    -2 & 0 & 0 &2 &
    8 & -6 & 0 & 2 & 0 & -2
    &&    220 & -12 & 4   &  0  & 0 & -6 &0 &    4
    \\
    4 &&
    660 & 4 & -6
 &    0 & -2 & 0 & 0 &
    -20 & 6 & -4& -2&4&0
    &&    560 & 16 & 2   &  0& -2 & 8 & -1 &    -4
    \\
    5 &&
    1408 & 0 & 4 
 &    -2 & 0 & 0 & 0 &
    32&4&0&-4&0&2
    &&    1144 & -24 & -8   & 4  & 0 & -12 & 0 &    4
    \\
    6 &&
    2794 & -6 & 4
 &    4 &0 & -2 & 0 &
   -30 &-8&2&0&2&0
   &&    2400 & 32 & 6   &   0&2 & 16 & 2 &    0
    \\
    7 &&
    5280 & 0 & -12 
 &     0 &0 & 0 & 0 &
    40&6&0&-2&0&0
    &&    4488 & -40 & 6   &   -2&2 & -20 & 0 &    -12
    \\
    8 &&
    9638 & 6 & 8 
 &   -2 &0 & -2 &2 &
    -58&2&-10&2&-2&2
    &&    8360 & 56 & -10   & 0&2 & 28 &0 &    8
    \\
    9 &&
    16960 & 0 & 4
 &    0 &0 & 0 &-2 &
    80&-14&0&2&0&0
    &&    14696& -72 & 8   & -4 &0 & -36 &0 &    8
    \\
    10 &&
    29018 & -6 & -16
 &  -2 &0 &2 & 0&
    -102&8&10&0&2&-2
    &&    25544& 88 & 2   & 4 &-2 &44 & 2&    -16
    \\
    11 &&
    48576 &0 & 12
 &    6 &0 &0 & 0&
    112&6&0&-2&0&2
    &&    42660 &-116& -18   &  0&-2 &-58 & 2&    12
    \\
    12 &&
    79530& 10& 6
 &    0& -2 &2 & 0&
    -150&-24&-6&0&2&0
    &&    70576 & 144& 16   &  -4& 0 &72 & 0&    4
    \\
    13 &&
    127776 & 0& -24 
 & -4& 0 &0 & 0&
    200&18&0&2&0&0
    &&    113520 & -176& 12   &0& 4 & -88& 0&    -24
    \\
    14 &&
    202050 & -14& 18
 &    0 & -2 &-2 & 2 &
    -230&12&10&4&2&0
    &&    180640 & 224& -26   &0   & 2 & 112&-2  &    16
    \\
    15 &&
    314688 &0& 12
&    -2 &0 &0& 0&
    272&-30&0&2&0&2
    &&    281808 &-272& 18  &8 &-2 &-136&-1 &    12
    \\
    16 &&
    483516&12& -36
&    6 & 0&0& 0&
    -348&24&-12&0&-4&2
    &&    435160 &328& 10&  0& -2&164& 0&    -32
    \\
    17 &&
    733920&0& 24
&  0& 0&0& 0&
    440&12&0&-4&0&0
    &&    661476 &-404& -42 &-4   & -2&-202&2 &    24
    \\
    18 &&
    1101364&-12&16 
& -6& 0&4& 0&
    -508&-44&20&-4&-4&2
    &&    996600&488&30 &  0   & 2&244& 0&    12
    \\
    \bottomrule
  \end{tabular}
}  }
  \caption{The number of  massive representations
    $A_g^{(a)}(n)$.
  }
  \label{tab:decompose_twisted}
\end{table}

\clearpage
\newpage

\end{document}